\documentclass{article}
\usepackage[utf8]{inputenc}
\usepackage{authblk}
\usepackage{hyperref}
\usepackage{bm}

\title {How One Quiet Man Became Everyone’s Sage: The Spiritual Recasting of Einstein}
\author{Galina Weinstein}
\affil{\normalsize The Department of Philosophy, University of Haifa, Israel.} 

\begin{document}

\maketitle

\begin{abstract}
This paper critically examines the central thesis of Kieran Fox’s \textit{I Am a Part of Infinity: The Spiritual Journey of Albert Einstein}—namely, that Einstein’s intellectual development constitutes a coherent spiritual path culminating in a form of pantheistic mysticism shaped by both Western and Eastern traditions. Fox presents Einstein as the modern heir to a long-suppressed lineage of rational spirituality, extending from Pythagoras and Spinoza to Vedanta and Buddhism, unified by wonder, reverence for nature, and a vision of cosmic unity. While Fox’s account is imaginatively rich and philosophically syncretic, it risks conflating distinct conceptual registers—scientific, metaphysical, and spiritual—and thereby oversimplifying Einstein’s intellectual complexity. Drawing on Einstein’s scientific writings and personal reflections, this study reconstructs a historically grounded portrait of his thought, emphasizing its tensions, ambiguities, and resistance to spiritual closure. The paper argues that Fox’s interpretation, though rhetorically compelling, substitutes a harmonizing spiritual mythology for the conceptual rigor and epistemic humility that defined Einstein’s actual worldview.
\end{abstract}

\tableofcontents

\section{Introduction}

The renewed interest in Albert Einstein’s reflections on religion and philosophy has given rise to a genre of interpretive literature that blends scientific biography with spiritual or metaphysical commentary. Kieran Fox’s \textit{I Am a Part of Infinity: The Spiritual Journey of Albert Einstein} is a recent and ambitious entry in this tradition. Framed as an exploration of Einstein’s “cosmic religious feeling,” the book presents a sweeping narrative in which Einstein emerges as the modern heir to a spiritual-philosophical lineage that includes Pythagoras, Spinoza, Vedanta, Buddhism, and Taoism. For Fox, Einstein’s science was not merely rational inquiry but a form of spiritual insight—an expression of cosmic awe, ethical nonviolence, and metaphysical unity.

Yet for all its rhetorical power and imaginative synthesis, Fox’s portrayal rests on interpretive moves that are conceptually fragile and often methodologically questionable. I show that Fox's thesis stands on what might be called scholarly chicken remains: a residue of fragmented sources loosely interpreted and elevated into an edifice of cosmic significance.

This paper is divided into three sections. Section \ref{2} reconstructs the main ideas of Fox’s book, focusing on its interpretive scaffolding and key conceptual claims. Section \ref{3} turns to Don Howard’s more disciplined account, “Einstein, ‘Cosmic Religion’, and Theology” \cite{How}, as a comparative benchmark for evaluating interpretive rigor and historical precision. Section \ref{4} evaluates Fox’s treatment of Einstein’s unified field theory, arguing that it imposes a metaphysical and quasi-spiritual narrative onto what was, for Einstein, a technical and methodologically rigorous program grounded in field-theoretic determinism, not mystical aspiration. Section \ref{5} addresses broader concerns regarding scholarly standards, particularly the use of Einstein’s writings, the dangers of philosophical retrojection, and the importance of distinguishing between poetic synthesis and historically responsible analysis.

\section{Einstein's Cosmic Religious Feeling} \label{2}

\subsection{Typology of Religious Evolution}

In his book \textit{I Am a Part of Infinity}, Kieran Fox offers an ambitious interpretation of Einstein’s “cosmic religious feeling” \cite{Ein54}, framing it not as a system of belief, dogma, or institutional faith but as the culmination of a three-phase historical and psychological development of religious consciousness. Drawing from Einstein’s 1930 essay "Religion and Science" \cite{Ein54}, Fox presents this schema as central to understanding Einstein’s secular spirituality, and he situates it in dialogue with Western scientific rationalism and Eastern philosophical traditions such as Vedanta, Buddhism, and Taoism.

Fox builds on Einstein’s typology of the three stages of religious development framework from the 1930 article “Religion and Science” \cite{Ein54}:

1. The religion of fear: Primitive responses to natural threats through the projection of supernatural agents.

2. Moral religion: Anthropomorphic theism organized around divine justice, reward, punishment, and providence (exemplified by historical world religions).

2. Cosmic religious feeling:
This highest and rarest phase is distinguished by the absence of dogma and anthropomorphic deity. It arises from a sense of awe at the “sublimity and marvelous order” of the cosmos, combined with a desire to experience the universe as a meaningful whole. It is independent of revelation or ritual and is found, Einstein notes, in individuals of exceptional insight—among them, he includes Democritus, Francis of Assisi, and Baruch Spinoza.

This third phase neither depends on revelation nor requires institutional mediation. It is available to all individuals, regardless of background, provided they cultivate mental discipline and openness to wonder.

\subsection{Relation to Eastern Spirituality}

Fox devotes considerable space to exploring the influence of Eastern traditions on Einstein’s thought. He traces Einstein’s exposure to Buddhist and Vedantic ideas through his reading of Schopenhauer, his travels in Asia, and the books in his personal library, which is part of the Einstein Archives. Fox argues that Einstein was sympathetic to non-dualism, ahimsa, and the ethical orientation of Eastern spirituality \cite{Fox}.

According to Fox, Einstein’s position resists traditional classification. While he rejected belief in a personal God, religious dogma, prayer, and the afterlife, he also criticized what he called “fanatical atheists” for failing to grasp the profound mystery and aesthetic sublimity of the universe.\footnote{Einstein to an unidentified addressee (1941) quoted in \cite{Jam}: "...there are the fanatical atheists whose intolerance is of the same kind as the intolerance of the religious fanatics and comes from the same source".} 
He described himself as a “deeply religious unbeliever” (as reported in the conversations with William Hermanns \cite{Her}). He maintained that science and cosmic religiosity were not enemies but mutually reinforcing expressions of the human spirit \cite{Fox}.

For Einstein, ethics did not require theological sanction. Moral behavior should arise from empathy, education, and social bonds, not from fear of divine punishment. But scientific work, particularly theoretical physics, required a near-religious devotion that, in his words, was inspired by cosmic religious feeling. This feeling, he claimed, gave scientists the strength to pursue lonely intellectual paths, often in the face of indifference or hostility.

Thus, Einstein’s spirituality is not reducible to atheism or traditional religion. It is best described as a rationalized reverence for the laws of nature and the human capacity to discern them—a form of secular transcendence that combines intellectual wonder with ethical seriousness \cite{Fox}.

\subsection{Wonder as the Sacred Pulse of Einstein’s Spirituality}

According to Fox, wonder is the core principle of Einstein’s spirituality, the wellspring of science and religion, and the foundation of “cosmic religious feeling.” The author intricately weaves wonder into Einstein’s spiritual and intellectual life by positioning it as a sacred impulse that replaces the traditional functions of faith, revelation, ritual, and dogma in prior religious systems.

Fox emphasized that Einstein deliberately decoupled spirituality from theology, scripture, or institutions, instead placing wonder at its center. Unlike earlier religions that are “petrified” in fixed beliefs or final answers, the cosmic religion is portrayed as open-ended, beginning not with certainty but with awe and questioning.
“Wonder represents a beginning, rather than an end; a call to adventure, rather than a conclusion” \cite{Fox}

Einstein’s spirituality is described as an invitation to explore mystery, not explain it away. In this sense, wonder is not the opposite of knowledge but the condition that makes authentic knowledge possible.

Fox intertwines emotion and intellect through Einstein’s concept of wonder. It is not merely sentimental reverence—it provokes inquiry and sustains curiosity. Wonder
“insists on our ignorance” – fostering humility;
“provokes honest inquiry” – fueling the pursuit of understanding; and “sustains our searching” – making scientific and philosophical investigation a spiritual task \cite{Fox}.
This positions wonder as epistemic humility, existential reverence, and the driving force of both Einstein’s personal development and humanity’s broader intellectual history.

A particularly significant move in Fox's book is the linkage of wonder not just to the universe outside, but to the mysterious power of the human mind itself. Einstein is described as being in awe of the mind’s ability to comprehend the cosmos:
“The most incomprehensible thing about the universe is that it is comprehensible.”
This paradox—our finite minds grasping the laws of an infinite cosmos—is, in Fox’s framing, one of the highest expressions of cosmic religious feeling. The idea that pure thought (Euclidean geometry, for instance) could yield certainty and clarity became a quasi-mystical experience of the mind’s potential for Einstein \cite{Fox}.

The quote, “The most incomprehensible thing about the universe is that it is comprehensible,” is one of the most frequently cited and paraphrased lines attributed to Einstein. It is derived from his 1936 essay "Physics and Reality," but people tend to cite it without qualification, giving it a more mystical flavor than Einstein intended. In Einstein’s usage, especially in "Physics and Reality," the quote is: "In speaking here of 'comprehensibility,' the expression is used in its most modest sense. It implies: the production of some order among sense impressions… The fact that it is comprehensible is a miracle" \cite{Ein54}.

Fox asserts that Einstein saw science as a spiritual activity driven by wonder. Far from tension with religion, scientific curiosity is framed as its highest modern incarnation (Fox quotes Einstein): “It is cosmic religious feeling that gives a man such strength... in this materialistic age of ours, the serious scientific workers are the only profoundly religious people” \cite{Ein54}.
This claim reverses traditional religious hierarchies: rather than priests or prophets, scientists and seekers of knowledge embody the deepest spiritual impulse because they live in a state of perpetual wonder \cite{Fox}.

Fox also notes that wonder must be paired with curiosity to avoid being manipulated or weaponized. Without intellectual integrity, awe can be dangerous—it can lead to idolizing leaders, ideologies, or nations (e.g., the cult of dictators). Einstein’s antidote to this distortion is critical wonder, or wonder guided by inquiry, rather than blind reverence.

Einstein’s reverence for wonder translates into a radically different vision of education and ethics:
Education should cultivate a “holy curiosity” (Hermanns conversations with Einstein \cite{Her}), not rote memorization.
Spiritual dignity derives from the unrestricted use of reason, not conformity to tradition.
A moral life emerges not from fear of punishment or desire for reward, but from awe at our place in the universe.

Thus, according to Fox, wonder is the pulse that animates Einstein’s worldview—scientific, ethical, and spiritual. It serves as a beginning to all serious inquiry; a substitute for traditional religious faith; a safeguard against nihilism and fanaticism; and ultimately, a universal access point to what Einstein called “cosmic religious feeling.”
Einstein’s cosmic religion, then, is not belief in a deity but a lived reverence for the universe's mystery and intelligibility. And at the heart of that reverence is wonder, not as a fleeting sentiment but as the foundation of a rigorous, responsible, and emotionally rich spirituality.

\subsection{The God of Spinoza and Influence of Schopenhauer’s Reading of Indian Philosophy}

Fox presents Einstein’s God as the God of Spinoza, but this “Spinozism” is suffused with Schopenhauerian and Indian philosophical themes, particularly through the lens of ahimsa, nonduality, and reverence for all life. 

The author frames Einstein’s view of God as aligned with Spinoza’s impersonal, non-anthropomorphic deity—a God identical with the lawful harmony of the universe, not a personal being who intervenes in human affairs: the cosmos as rational, ordered, and sacred; the idea that ethical life flows from an intuitive grasp of unity; the rejection of dogma, miracles, and divine commands; and the emphasis on the natural world and reason as the only true sources of spiritual insight. “As man becomes conscious of stupendous laws that govern the universe in perfect harmony, he begins to realize how small he is. ... This is the beginning of cosmic religion within him” (Hermanns conversation with Einstein \cite{Her}).
This echoes Spinoza’s notion that understanding nature is a kind of salvation, and that ethical action follows naturally from an intellectual love of God/Nature (\emph{amor Dei intellectualis}). Einstein, like Spinoza, sees divinity not above or outside the world, but immanent within it \cite{Fox}.

Fox strongly implies that Einstein’s ethical and spiritual orientation—his emphasis on ahimsa and cosmic unity—is shaped by Schopenhauer’s reading of Indian philosophy, even if Einstein absorbed it indirectly.
He explicitly quotes Schopenhauer:
“All genuine virtue proceeds from the immediate and intuitive knowledge of the metaphysical identity of all beings.”
This metaphysical insight—that all beings are One—is portrayed as the foundation of Einstein’s ethics, especially his pacifism and reverence for life.
The author traces Einstein’s vegetarianism and compassion for animals not to the Bible or Judaism, but to Eastern traditions—especially Jainism, Buddhism, and the Upanishads.

Schopenhauer is presented as the gateway through which these ideas entered Einstein’s worldview. Schopenhauer called Indian texts “the fruit of the highest human knowledge and wisdom.” He admired the Upanishads, praised Brahmanic nondualism, and advocated for vegetarianism as moral progress.
Fox highlights how Einstein absorbed Schopenhauer’s reverence for Eastern ethics, even if Einstein was not a systematic student of Indian philosophy himself.

Einstein’s ethical ideals are also reflected in his admiration for Mahatma Gandhi and Albert Schweitzer, who espoused and lived by principles closely aligned with ahimsa.
Gandhi’s Jain-inspired nonviolence is a real-world enactment of the metaphysical unity that Einstein and Schopenhauer philosophized about \cite{Fox}.

\subsection{Hindu Elements: Nonduality (Advaita), Ahimsa, and Spiritual Self-Realization}

Fox then emphasizes the oneness of all existence, echoing Advaita Vedanta: “All things, including even our minds, were woven together in a seamless, self-consistent tapestry.”
He again reiterates Einstein’s growing appreciation for ahimsa, or nonviolence rooted in spiritual awareness, through his admiration for Indian sages and the broader Indian tradition \cite{Fox}.

Einstein’s spiritual cosmology, as presented by Fox, forms the metaphysical foundation for a deeply ethical humanism and a universalist political sensibility. The insistence that freedom is not a given but a rare attainment—requiring sustained inner work and detachment from egocentric instincts—aligns with Indian philosophical traditions, particularly those of Vedanta and Jainism, where liberation (moksha) is achieved only through disciplined self-overcoming and moral clarity. In this framework, freedom becomes a spiritual task, not an inherited right, and its cultivation entails ethical responsibility toward others. Fox quotes Einstein: “Only if outward and inner freedom are constantly and consciously pursued,” Einstein wrote, “is there a possibility of spiritual development” \cite{Ein54}. He interprets such statements as underscoring Einstein's belief that autonomy must be actively earned through inner transformation and moral striving \cite{Fox}.

At the metaphysical level, this conception of freedom is embedded within a Spinozist determinism, where all individuality and autonomy are ultimately illusory—“a part of the whole” \cite{Cal}, as Einstein put it. Fox cites Spinoza’s metaphor of the "tiny worm living in the blood," which reinforces this non-dual ontology, where the finite self cannot fully perceive its embeddedness within the infinite totality. In this view, Einstein’s God, like Spinoza’s, is impersonal and immanent, a lawful harmony rather than a providential agent. “My God is too universal to concern himself with the intentions of every human being” (Hermanns conversations with Einstein \cite{Her}), Einstein remarked, distancing himself from theistic personalism and emphasizing the divine as cosmic order rather than moral judge \cite{Fox}.

Einstein’s determinism is also influenced by Schopenhauerian pessimism, particularly in his emphasis on unconscious drives and physiological determinants of action. Echoing Schopenhauer’s doctrine of the metaphysical Will, Einstein admitted that the scope of human freedom is constrained by inner compulsions beyond the reach of reason: “We can do what we wish, but we can only wish what we must” (Einstein's conversations with George Sylvester Viereck \cite{Vie}). Here, ethics is no longer predicated on rational choice but on the more profound intuition of unity—a realization that all beings are expressions of the same underlying reality. As Schopenhauer and Einstein suggest, ethical conduct—especially compassion, pacifism, and reverence for life—arises not from duty imposed by external commandments, but from a direct metaphysical awareness of interdependence \cite{Fox}.

This metaphysical orientation carries clear political consequences. Einstein’s calls for pacifism, social justice, and cosmopolitan unity were not grounded in abstract political theory, but in a spiritualized vision of the human being as a node in an interwoven totality. His ethical universalism was a direct outgrowth of his spiritual intuition of non-separateness. As such, his rejection of nationalism, militarism, and racial hierarchy can be read as political judgments and metaphysical imperatives. “Humanity,” he wrote, “is one and undivided” (Hermanns conversation with Einstein). Ethical decisions must be evaluated not through tribal or sectarian lenses, but through the lens of cosmic interconnectedness \cite{Fox}.

\subsection{Einstein’s commitment to unity: Jnana marga of the West}

Fox argues that Einstein’s spiritual orientation shares striking affinities with Hindu, Jain, and Buddhist metaphysical and ethical frameworks. In particular, he explores how Einstein’s commitment to unity, nonviolence, and ego-transcendence echoes the principles of Advaita Vedanta, ahimsa, and the jñāna mārga (path of knowledge) \cite{Fox}.

Einstein’s belief in a deterministic, lawful cosmos—where individuality is an “optical delusion of his consciousness” \cite{Cal}—is interpreted as a form of modern nonduality, paralleling the Vedantic doctrine of Brahman as the sole reality behind the illusory multiplicity of the world (Māyā). His ethical universalism, commitment to pacifism, vegetarianism, and reverence for life are aligned with Jain and Buddhist moral ideals. Drawing on Einstein’s references to Spinoza and Schopenhauer, Fox traces a philosophical lineage that unites Spinozist pantheism, Schopenhauerian metaphysical Will, and Indian monism under a shared vision of inner liberation through insight into the interconnectedness of all things \cite{Fox}.

The author further argues that Einstein’s scientific ethos—his pursuit of unification through unified field theory and his belief in the rational intelligibility of the cosmos—was undergirded by this spiritual disposition. This cosmic religiosity, which Einstein associated with humility and the transcendence of ego, is seen as a moral and epistemological orientation that bridges the divide between scientific rationalism and spiritual insight. Fox invokes Einstein’s admiration for Gandhi, his affinity for Taoist and Upanishadic texts, and his espousal of the Spinozist \emph{amor Dei intellectualis} to support the claim that Einstein’s intellectual path exemplified the ancient ideal of enlightenment through knowledge—a modern, secular expression of the jñāna mārga \cite{Fox}.

\subsection{Relation to the Western Tradition}

Fox presents a revisionist account of the Western intellectual tradition, reinterpreting it not as the product of Judeo-Christian theism but as the outgrowth of a suppressed lineage of Pythagorean-Platonic pantheism. He explicitly rejects "the dependency thesis", the view that “without Christianity, modern science could not exist”— arguing instead that institutional Christianity stifled scientific inquiry during the Middle Ages. He situates the true spiritual roots of modern science in ancient Greek rationalism, tracing a lineage from Pythagoras through Plato, Neoplatonism, the Renaissance, and early modern physics to Einstein, whom he portrays as the culmination of this tradition. 

In Fox’s account, Einstein is not merely a physicist but the modern heir to a Western current that fused rational inquiry with spiritual reverence for mathematical harmony. The hallmark of this alternative genealogy is a spirituality rooted not in revelation or dogma, but in the contemplative apprehension of cosmic order. Ultimately, Fox romanticizes the West as a spiritually charged intellectual tradition that is neither theistic nor materialist but grounded in mathematical pantheism, with Einstein as its most dazzling expression \cite{Fox}.

\subsection{The Pantheistic Lineage: From Pythagoras to Einstein}

Fox constructs an interpretive genealogy that links Pythagoreanism, Giordano Bruno, Baruch Spinoza, and Einstein into a single metaphysical lineage. His strategy is literary, mainly metaphorical. He draws on symbolic continuities and thematic resonances to craft a pantheistic tradition culminating in Einstein’s “cosmic religion.” The result is a spiritualized intellectual history \cite{Fox}:

\emph{1. Pythagoreanism as proto-pantheism:} Fox opens this lineage by invoking Pythagoreanism as the primordial source of a worldview in which mathematics, harmony, and metaphysical unity are foundational. 

\emph{2. Giordano Bruno as the radical bridge:} Giordano Bruno functions in Fox’s narrative as the transitional figure who revives and reconfigures Pythagorean themes in early modern Europe. Fox presents Bruno as a heretical mystic who asserted the infinite nature of the cosmos and the divinity of all matter. Bruno’s pantheism—articulated through poetic evocations of an immanent Infinite—is dramatized as an act of spiritual rebellion that anticipates Spinoza.
Fox emphasizes Bruno’s persecution and eventual execution as emblematic of the cultural resistance to pantheistic metaphysics. 

\emph{3. Spinoza as the pantheist philosopher par excellence:} At the center of Fox’s pantheistic lineage stands Baruch Spinoza, whom he portrays as the most precise and influential philosopher of divine immanence in the Western tradition. 

\emph{4. Einstein as the modern heir:} In Fox’s account, Einstein emerges as the culminating figure in this pantheistic succession. Drawing on Einstein’s frequent references to “Spinoza’s God " and his notion of “cosmic religious feeling,” Fox interprets Einstein as the modern embodiment of Pythagorean harmony, Bruno’s mystical audacity, and Spinoza’s rationalist metaphysics.

In this framing, Einstein is cast not merely as a physicist but as a sage: like Bruno, a heretical prophet of unity; like Spinoza, a rigorous thinker of divine immanence; and like Pythagoras, a contemplative of cosmic order through number. Fox reads Einstein’s awe before the laws of nature as a spiritual stance, and implicitly equates this awe with the mystical insight sought by earlier pantheists.

\subsection{Einstein as a Modern Pythagorean}

Fox constructs a sweeping intellectual genealogy, positioning Einstein as the culmination of the Pythagorean tradition. His central claim is that Einstein’s scientific worldview was not merely rational or empirical but deeply rooted in a philosophical-religious sensibility that originated with Pythagoras. The foundation of this claim rests on several thematic pillars \cite{Fox}:

\textit{Faith in mathematical harmony}: Fox identifies Einstein’s conviction in the mathematical intelligibility of the universe as a defining trait linking him to Pythagoreanism. Einstein’s oft-cited belief that “pure thought can grasp reality” \cite{Ein54} is evidence of a metaphysical commitment to mathematical simplicity and unity—a conviction shared by the Pythagoreans, who viewed number and proportion as the underlying principles of all existence.

\textit{The role of aesthetic and spiritual intuition in Science}: Fox emphasizes Einstein’s statements about imagination, intuition, and the emotional or ecstatic experience of scientific discovery. He interprets Einstein’s descriptions of “rapture” \cite{Isa} and “cosmic religious feeling” \cite{Ein54} as spiritual episodes analogous to the mystical insights sought by the Pythagoreans through ascetic and contemplative disciplines.

\textit{Philosophical continuity from antiquity to modern science}: By tracing intellectual influences from Pythagoras to Plato, through the Neoplatonists and early modern scientists like Kepler and Newton, Fox situates Einstein as the heir of a long tradition of thinkers who combined rational inquiry with metaphysical aspiration. This tradition, he argues, was not derivative of Judeo-Christian theology (as claimed by proponents of the “dependency thesis”) but of ancient Greek natural philosophy.

\textit{Einstein’s admiration for Democritus and the pre-Socratics}: Drawing on Einstein’s personal library (especially his annotated copy of Maurice Solovine’s book on Democritus), Fox argues that Einstein recognized and revered the philosophical lineage descending from the Pythagoreans through Democritus. Solovine explicitly links Democritus to Pythagorean teachings, reinforcing the sense that Einstein saw himself as part of this continuum.

\textit{Spiritual reverence without Theism}: Fox argues that Einstein’s cosmic religiosity was not mystical in the theistic or supernatural sense, but Pythagorean in that the divine was immanent in the rational order of nature. Einstein’s rejection of a personal God and his embrace of Spinoza’s naturalistic determinism are modern expressions of Pythagorean pantheism.

\textit{Scientific imagination as modern mysticism}:
Fox further aligns Einstein with the Pythagorean tradition by likening Einstein’s use of thought experiments to ancient visionary experiences. While Einstein insisted he was not a mystic, Fox contends that his imaginative approach to physics mirrored the ancient practice of accessing metaphysical truths through spiritual exercises.

\textit{Testimony from other physicists}:
The case is bolstered with reflections from Erwin Schrödinger, Werner Heisenberg, Wolfgang Pauli, and Bertrand Russell, all of whom, as Fox shows, acknowledged the Pythagorean character of modern physics. Their admiration for the tradition reinforces Fox’s broader thesis that Einstein did not stand outside but rather at the apex of this intellectual and spiritual arc.

\textit{Musical metaphors}: Fox argues that Einstein’s musical metaphors and preference for harmonic causality over linear determinism reveal the influence of Pythagorean cosmology, in which the universe's structure is conceived as an interlocking system of resonances. The remark—"we are like a juvenile learner at the piano just relating one note to that which immediately precedes or follows” \cite{Pla},\footnote{“We are like a child who judges a poem by the rhyme and knows nothing of the rhythmic pattern. Or we are like a juvenile learner at the piano just relating one note to that which immediately precedes or follows” \cite{Pla}. This is a problematic source use, as discussed in the next section.} suggests epistemic humility and an ethical imperative: to learn how to attune ourselves to the deeper rhythms of the cosmos. This vision seamlessly fuses Einstein's scientific rationalism with spiritual insight, yielding what may best be termed a secular cosmic mysticism.

Fox’s argument is interpretive and synthetic, weaving historical, philosophical, and psychological threads into a narrative that frames Einstein as the “pinnacle” of a tradition that merges science with sacred inquiry. While Einstein seldom referred explicitly to Pythagoras, Fox maintains that his philosophical assumptions—mathematical Platonism, epistemic humility, and aesthetic faith in the unity of nature—align closely with those of the ancient school. Hence, Einstein becomes not just a physicist of the modern age, but the most dazzling expression of an ancient dream: to know nature through number is to touch the divine.

\section{Reframing Einstein’s Cosmic Religious Feeling} \label{3}

While Fox offers an earnest and often lyrical interpretation of the spiritual dimensions of Einstein’s worldview, this paper argues that it substitutes poetic resonance for historical and conceptual rigor, resulting in a portrait that obscures the complexity and development of Einstein’s thought. In Fox’s hands, Einstein emerges as a Spinozist pantheist, a modern Pythagorean, a philosophical heir to Schopenhauer, an admirer of Gandhi, and a dialogical companion to sages of the Upanishads and the Tao Te Ching—all harmonized under the expansive canopy of “cosmic religious feeling.” Though these associations find selective support in Einstein’s writings or intellectual sympathies, their fusion into a unified spiritual archetype may reflect more of Fox’s syncretism than Einstein’s self-conception.

Einstein’s tone, particularly in his reflections on religion and philosophy, was marked not by mystical exuberance but by restraint, irony, and intellectual humility. He eschewed metaphysical speculation for what he called, in his 1936 essay “Physics and Reality,” the “most modest sense” of comprehensibility—the ability to impose some conceptual order on sense impressions \cite{Ein54}. That this was even possible, he called a “miracle,” not in the theological sense, but as an expression of astonished reason. In a letter to a sixth-grade student, Phyllis Wright, written the same year, Einstein explained that although science rests on the assumption of lawful regularity, this conviction too is a kind of belief—one tempered by the awareness of our epistemic limitations.\footnote{\href{https://www.sothebys.com/en/buy/auction/2023/fine-books-and-manuscripts/einstein-albert-a-remarkable-letter-to-phyllis?locale=de}{Einstein’s 1936 letter to Phyllis Wright, Sotheby’s Fine Books and Manuscripts Auction.}}

Were Einstein alive to read Fox’s spiritual kaleidoscope, he might have responded not with indignation but wry detachment—a bemused smile rather than a metaphysical affirmation. He was not profoundly religious but filled with awe. In honoring the vastness of Einstein’s thought, one must also honor his wariness of theological constructs, his rejection of anthropomorphic deities, and his reluctance to align with spiritual traditions, Western or Eastern. His was a kind of spiritual naturalism—reverent but unsentimental, disenchanted yet luminous.

Although sincerely engaged and occasionally illuminating, Fox’s interpretation tends to elevate metaphor into metaphysics and inspiration into identification, attributing to Einstein a doctrinal coherence he explicitly disavowed. The result is a spiritually evocative but historically fragile reconstruction—that Einstein, with his characteristic iconoclasm, might have met with polite interest before returning to his physics.

\subsection{Don Howard: Einstein’s Cosmic Religion and God}

In his comprehensive study, "Einstein, 'Cosmic Religion,' and Theology", Don Howard presents Einstein’s cosmic religion not as a traditional faith but as a deeply intellectual and emotionally resonant spiritual orientation. Rather than affirming a personal deity or embracing institutional religion, Einstein articulated a vision of religiosity rooted in awe before the rational structure of the universe. This “cosmic religious feeling,” as he termed it, reflects a yearning to transcend conventional religion's personal and moralistic aspects in favor of a profound engagement with what he called “reason made manifest in nature.” For Einstein, this was not merely metaphorical language: it expressed a conviction that the universe operates according to an underlying rationality, accessible to the human mind through the combined efforts of science and contemplation \cite{How}.

While Einstein is often associated with Spinoza, whose concept of God as nature resonated with his impersonal theology, Howard argues that this identification is misleading and overly simplistic. Spinoza’s metaphysical monism, which dissolves the reality of individual things into the substance of God, contrasts with Einstein’s strong emphasis on the separability and individuation of events in space-time. Instead, Howard suggests that Einstein’s conception of divinity is better understood through the ancient Stoic and Philonian idea of logos—a rational principle pervading the cosmos. Logos, for both the Stoics and Philo of Alexandria, is not a personal god but a structuring force that makes the universe law-governed and intelligible. Einstein’s repeated references to the “reason revealed in nature” align more closely with this Hellenistic tradition than Spinoza’s pantheism \cite{How}.

At the same time, Howard highlights that Einstein’s spiritual vision contains elements drawn from other traditions. Influenced by Schopenhauer, Einstein viewed aesthetic experiences—especially musical ones—as windows into the hidden rational order behind appearances, akin to glimpses through what Schopenhauer, borrowing from Vedantic Hinduism, called the “veil of Maya.” While Einstein himself did not use this metaphor explicitly, Howard argues that the structure of Einstein’s thought mirrors this conception: science and music, in their highest forms, allow for epiphanic encounters with the underlying order of existence. In this sense, Einstein’s spirituality can be seen as Schopenhauerian, not in metaphysics per se, but in its emphasis on intellectual and aesthetic transcendence of the individual self \cite{How}.

Howard also notes that Einstein admired elements of Eastern thought, notably Buddhism, which he saw as embodying a form of religious feeling devoid of dogma and centered on ethical and cosmic awareness. However, Howard cautions against interpreting Einstein as a mystic in any traditional Eastern or monastic sense. Instead, Einstein’s spirituality is defined by its grounding in rationality, detachment from anthropomorphic theology, and deep moral seriousness, without invoking supernaturalism or salvation.

Crucially, Howard argues that Einstein did not separate science and religion, but neither did he conflate them. Instead, he saw both as springing from the same source: a profound emotional and intellectual response to the mystery and order of the universe. For Einstein, science is a sacred endeavor, motivated by cosmic religious feeling, and religion, at its most refined, is the emotional reverberation of a mind striving to understand nature’s rational structure. Thus, Einstein envisions a unity of science and spirituality—not through theology, but through a reverent, epiphanic recognition of the intelligibility of the cosmos \cite{How}.

Howard’s reading of Einstein presents a rich and nuanced picture: Einstein’s cosmic religion emerges as a unique synthesis of Stoic logos, Schopenhauerian aestheticism, selective Eastern influences, and scientific rationalism. It is a spirituality without dogma, a reverence without theism, and a religious feeling anchored in the intellectual and emotional pursuit of understanding the universe.

Einstein's pursuit of a unified field theory became central to his scientific identity in his later years, just as cosmic religion became central to his philosophical and emotional worldview. While Howard does not name it in his paper, the unified field theory is a concrete instantiation of Einstein's cosmic religion. In Einstein’s eyes, the unified field theory was not just a technical project, but a metaphysical affirmation of that logos: the rational intelligibility of the cosmos in its most complete form. Howard stresses that Einstein saw mathematical unity and simplicity as signs of truth. This very conviction drove the unified field theory. So while the unified field theory is absent from the essay, its spirit permeates Howard’s portrayal of Einstein.

\subsection{From Intellectual Modesty to Secular Mysticism}

Einstein’s best-known articulation of “cosmic religious feeling” appears in his 1930 essay "Religion and Science." There, he characterizes this sentiment as a non-theistic, non-anthropomorphic response to the perceived order and intelligibility of the universe. He is explicit in distancing it from traditional religious belief \cite{Ein54}:

\begin{quote}
There is a third stage of religious experience... I shall call it cosmic religious feeling... There is no anthropomorphic conception of God corresponding to it... The individual feels the futility of human desires and aims and the sublimity and marvelous order which reveal themselves both in nature and the world of thought. 
\end{quote}

Einstein’s language here is contemplative and restrained, marked by epistemic humility rather than metaphysical declaration. His tone reflects a naturalistic worldview, grounded in emotional engagement with the structure of reality but unencumbered by supernatural claims.

But Fox’s account tends to blur these distinctions by introducing metaphors and imagery not native to Einstein’s writings drawn from Schopenhauer’s interpretation of Indian philosophy, like the notion that we “dance to a mysterious tune,” which originates not with Einstein himself but with William Hermanns’ retrospective rendering in \textit{Einstein and the Poet} \cite{Her, Fox}.\footnote{The full quote from Hermanns reads:
\begin{quote}
He [Einstein] pointed towards the window: “If we look at this tree outside whose roots search beneath the pavement for water, or a flower which sends its sweet smell to the pollinating bees, or even our selves and the inner forces that drive us to act, we can see that we all dance to a mysterious tune, and the piper who plays this melody from an inscrutable distance—whatever name we give him—Creative Force, or God—escapes all book knowledge.” \cite{Her}
\end{quote}} While such treatments may offer affective resonance or imaginative appeal, they tend to displace Einstein’s rigorously naturalistic perspective with a more impressionistic spirituality that lacks historical grounding.

While Einstein cited Schopenhauer approvingly \cite{Ein54}, he did not adopt his metaphysical pessimism or the Vedantic substratum that underpins Schopenhauer’s notion of the Will. Thus, attributing Schopenhauerian or Vedantic metaphysics to Einstein risks overstatement and misrepresents his approach's philosophical restraint.
Still, as Howard emphasizes in his study, Schopenhauer’s influence on Einstein should not be dismissed entirely. Schopenhauer shaped Einstein’s views on determinism, particularly through the maxim “A man can do what he wants, but not want what he wants,” which Einstein often cited as a source of intellectual consolation. Moreover, Schopenhauer’s notion that music grants a glimpse beyond the veil of Maya, offering access to the “innermost kernel of things,” resonates with Einstein’s reflections on music as a mode of nonverbal, intuitive thinking. 

Howard suggests that this epiphanic function of music—its ability to momentarily lift one above the world of the personal and into contact with a deeper order—parallels the structure of Einstein’s cosmic religious feeling, where the impersonal pursuit of reason in nature becomes a kind of spiritual act. Thus, while Einstein rejected Schopenhauer’s metaphysical Will, he retained elements of his epistemology and aesthetic philosophy—particularly the role of music as a non-discursive mode of insight—which helped shape his unique form of secular reverence grounded in reason and wonder \cite{How}.

Einstein was scrupulous in distinguishing between metaphor (e.g., about music, God, or wonder) and metaphysics, and his published writings consistently attempt to separate poetic language from ontological commitment. When secondary or anecdotal sources are introduced without contextual framing, they risk projecting a tone of cosmic mysticism that Einstein neither endorsed nor encouraged. 
What emerges in Fox's text, then, is not the historically grounded Einstein who spoke with restraint and precision, but a stylized figure shaped by contemporary spiritual discourse. This interpretive overlay, however eloquent, substitutes metaphorical amplification for historical fidelity and aesthetic inspiration for analytical clarity. Doing so, it obscures the distinctive combination of rigor, modesty, and conceptual discipline that characterized Einstein’s worldview.

\subsection{Thematic Conflation and Lack of Conceptual Distinctions}

A further methodological concern in Fox’s account lies in its thematic conflation—specifically, its failure to distinguish between Einstein’s scientific reasoning, philosophical reflections, and personal or spiritual sensibilities. These domains, which Einstein himself carefully demarcated, are interpreted through a single unifying lens of mystical wonder. The result is a flattened portrayal of a thinker whose intellectual life was marked by conceptual rigor, disciplinary clarity, and a deep commitment to epistemological precision. 

Howard shows how Schopenhauer’s aesthetic epistemology influenced Einstein’s understanding of music and insight. He demonstrates how cosmic religious feeling motivated scientific work, but never claims these were equivalent modes of thought. He does not suggest that scientific breakthroughs are spiritual insights, only that they are motivated by a spiritualized feeling of awe at reason \cite{How}. By contrast, Fox collapses distinctions that Howard takes great care to preserve while still contextualizing.

Fox constructs a seamless arc of spiritual continuity from Einstein’s childhood wonder to special relativity and cosmic religion. Consider, for instance, Einstein’s operational definition of simultaneity in special relativity—a breakthrough grounded in methodological innovation and logical construction. In Fox’s narrative, such developments are rendered with spiritual overtones, as though they were moments of intuitive revelation rather than the outcome of sustained analytical effort. This reframing abstracts Einstein’s work from its technical and historical contexts and repositions it within a metaphysical narrative he did not embrace.

Similarly, Einstein’s reflections on the harmony of natural law, though often expressed with aesthetic admiration, remain firmly within a rationalist tradition. His writings—particularly those collected in \textit{Ideas and Opinions} \cite{Ein54}—consistently affirm a naturalistic worldview. While he acknowledged the emotional resonance of the mysterious and the sublime, he did so without recourse to mystical language or spiritual doctrine. His sense of wonder was grounded in intelligibility, not transcendence.

Fox’s narrative, however, tends to collapse these distinctions. Drawing unqualified connections between Einstein’s thoughts on magnetism, simultaneity, and “cosmic religious feeling” constructs a seamless arc of spiritual continuity. This approach dissolves the epistemic boundaries that Einstein maintained between physical theory, philosophical reflection, and personal sentiment.

While such narrative synthesis may serve a broader inspirational aim, it comes at the cost of analytical integrity. It overlooks that Einstein’s intellectual practice was built upon differentiating forms of inquiry and recognizing the distinct modes of reasoning appropriate to each. Physical theories were to be logically constructed and empirically tested; philosophical reflections were carefully reasoned; and emotional responses, though never disavowed, were not mistaken for epistemic claims.

Such portrayals risk misrepresenting Einstein's content and architecture by collapsing these layers into a generalized spiritualism. To appreciate the depth of his intellectual contribution, what is needed is not interpretive fusion but careful reconstruction—one that preserves the disciplinary distinctions he was at pains to maintain.

\subsection{The Popular Recasting of Einstein’s Intellectual Persona}

Popular portrayals of Einstein often downplay the irony, restraint, and epistemic caution that characterize his reflections on science, religion, and metaphysics. His writings on these subjects were seldom systematic or doctrinaire; instead, they reveal a sensibility marked by provisional reasoning, strategic ambiguity, and an acute awareness of the conceptual limits of knowledge.

Einstein frequently employed metaphors, understatement, and hedged formulations to resist simplistic classifications, positioning himself neither within conventional religiosity nor among militant atheists. His references to “God” were rhetorical devices expressing philosophical commitments to order and intelligibility, not theological affirmations. The famous remark “God does not play dice,” often misunderstood as metaphysical, was a critique of quantum indeterminacy, rooted in his preference for lawful regularity.

Fox’s narrative, however, recasts Einstein in a consistently elevated and emotionally univocal tone. It smooths out the dialectical structure of his thinking—the ongoing tension between explanation and mystery—and replaces it with a narrative of spiritual fulfillment and cosmic attunement. In doing so, it obscures Einstein’s intellectual modesty. It presents a persona more aligned with the archetype of a sage than the historically grounded figure found in his correspondence, scientific writings, and philosophical essays.

This interpretive shift becomes especially evident in the treatment of Einstein’s Spinozist conception of God. While Fox acknowledges Einstein’s public avowal of “Spinoza’s God,” he overlays it with language suggesting a more mystical or quasi-theistic outlook.
Einstein is presented simultaneously as a Spinozist and a spiritual humanist, a naturalist and a visionary—an interpretive fusion that elides key distinctions. Yet Einstein was explicit in drawing boundaries between metaphor and metaphysics. 

His 1929 telegram to Rabbi Herbert S. Goldstein remains one of the most precise articulations of his worldview \cite{CPAE16}, Doc. 508:

\begin{quote}
I believe in Spinoza’s God, who reveals himself in the orderly harmony of what exists, not in a God who concerns himself with human beings' fates and actions. 
\end{quote}

This statement is precise, public, and philosophically restrained. Einstein dissociates himself from personal or providential conceptions of God and reaffirms a non-anthropomorphic commitment to rational order. His “cosmic religious feeling” is not spiritual pantheism but a form of epistemic reverence—a deeply felt response to the coherence and intelligibility of the universe, unaccompanied by metaphysical or devotional content.

By portraying intellectual curiosity as a spiritual vocation and blending Einstein’s rhetorical gestures with quasi-mystical language, Fox projects a metaphysical identity onto Einstein that he did not claim for himself. The phrase “curiosity was therefore the sine qua non of the sincere spiritual seeker” \cite{Fox}, for instance, attributes to Einstein a fusion of scientific inquiry with spiritual aspiration that lacks sufficient support in the textual and historical record.

Moreover, the context of Einstein’s 1929 statement \cite{CPAE16}, Doc. 508, crafted in response to theological misreadings of relativity, underscores its strategic and clarificatory nature. It was not a spontaneous declaration of spiritual belief but a deliberately measured intervention in a public controversy. The reference to “Spinoza’s God” functioned as rhetorical shorthand for Einstein’s commitment to natural law, not as an endorsement of mystical speculation.

Fox’s substitution of a spiritualized Einstein for the historically situated one risks flattening the nuance of Einstein’s thought. What is lost is not only fidelity to his philosophical stance but also the opportunity to appreciate the integrity of a worldview built upon conceptual discipline, emotional restraint, and deep ethical and epistemic commitments.

\subsection{Einstein’s God: Metaphor, Irony, and the Order of Nature}

To speak of “God” in Einstein’s writings is to enter a rhetorical domain shaped by metaphor, irony, and philosophical nuance. As Leopold Infeld once observed, Einstein used the term “more often than a Catholic priest” \cite{Inf}—but never in a conventionally theological sense. 
This metaphorical usage appears early in Einstein’s correspondence. In 1901, he quipped to Marcel Grossmann, “God created the donkey and gave him a thick skin” \cite{CPAE1}, Doc. 100. And in 1905, shortly after formulating the mass–energy relation, he wrote to Conrad Habicht, “God Almighty might be laughing at the whole matter and might have been leading me around by the nose” \cite{CPAE5}, Doc. 28. Such remarks are not expressions of belief, but instances of ironic modesty—deploying “God” as a rhetorical figure to signal uncertainty and intellectual humility.

Even in more serious contexts, Einstein retained this tone. One of his most cited aphorisms, engraved at Princeton’s Fine Hall, reads: “The Lord is subtle, but malicious he is not” \cite{Pais}. 
The remark reportedly responded to Dayton Miller’s ether-drift experiments, which seemed to challenge special relativity. While some physicists were alarmed, Einstein held that ambiguous data should not displace a well-corroborated theory. Here, “God” does not refer to a divine agent but to nature’s intelligible structure—subtle, yet ultimately lawful and comprehensible.

Einstein’s use of religious idiom frequently served aesthetic or epistemological functions. Remarks such as “God does not care about our mathematical difficulties; He integrates empirically” \cite{Inf} are conceptual evaluations, not theological claims. “God” symbolizes the rational coherence he sought in nature, not a supernatural entity.

His well-known critique of quantum indeterminacy—“God does not play dice”—is among the most misunderstood. It articulates his philosophical resistance to probabilistic accounts of physical law, not an article of religious conviction. In a later exchange with John Wheeler, Einstein reflected, “Well, I still can’t believe God plays dice… but maybe I’ve earned the right to make my mistakes” \cite{Whe}, revealing irony rather than dogmatism.

As a whole, Einstein’s recurrent invocations of “God” are best read as metaphorical gestures—rhetorical devices expressing awe at nature’s intelligibility and reverence for rational inquiry. They reflect a worldview committed to logical coherence, empirical rigor, and aesthetic clarity. Far from affirming mysticism or theism, his idioms borrow the language of religion to articulate ethical and epistemic ideals embedded in science.

\subsection{Significance of Wonder for Einstein}

In his later reflections, Einstein described wonder not as a transient emotional response, but as a profound and sustained mode of consciousness—a kind of existential astonishment that bound together unconscious intuition and rational inquiry. For him, wonder was the origin of the scientific impulse and its enduring companion: a deep affective response to the mysterious intelligibility of nature, which animated and sustained the search for understanding \cite{How}.

Einstein often portrayed wonder as a reverent perplexity—a spontaneous disruption of habitual thought in the presence of something inexplicably ordered. It was, in his words, “the fundamental emotion that stands at the cradle of true art and true science,” a response not only to ignorance or surprise but to the “profoundest reason and the most radiant beauty” \cite{Ein54} only dimly accessible to the human mind. This experience of mystery, Einstein held, constituted the essence of what he called “cosmic religious feeling”—a dogma-free, impersonal sense of awe before the rational structure of the universe.

Among his retrospective reflections, Einstein identified his childhood encounter with a compass, at the age of four or five, as his earliest experience of wonder. The compass needle’s unwavering orientation—seemingly immune to visible contact or mechanical causation—conflicted with his naïve assumptions about how the world worked. This paradox of unseen agency left what he later described as a “deep and lasting impression,” awakening a lifelong curiosity about invisible forces and the intelligibility of nature \cite{Ein49a}.

Yet this account must be carefully contextualized. Composed in his \emph{Autobiographical Notes} (1949), when Einstein was seventy, the story is not a raw recollection but a retrospective construction, shaped by his mature philosophical sensibilities. The compass scene should be read not as a transcript of childhood cognition, but as an intellectualized memory—a moment reinterpreted through the lens of an identity formed by decades of scientific and metaphysical reflection. To understand Einstein’s wonder is therefore to hear not the child’s voice, but the adult's carefully crafted narrative: one in which wonder becomes not just a beginning, but a principle of continuity in the life of the mind.

Einstein repeatedly affirmed the importance of retaining childlike openness to wonder as essential to scientific creativity. He believed that the rare ability to ask simple, naïve questions—questions usually abandoned with age—had led him to fundamental problems in space and time. These were, he suggested, the kinds of questions that “only children continue to ask” into maturity \cite{Hol}.

In his Autobiographical Notes, Einstein explained that “thinking goes on for the most part without use of signs (words),” and arises largely from unconscious processes \cite{Ein49a}. Wonder, for him, preceded articulation: it was pre-verbal, intuitive, and immersive, a mode of direct experience rather than discursive reasoning. His memory of the compass illustrates this vividly—it was not processed through analysis, but felt as a silent dissonance that summoned inquiry.

At twelve, Einstein encountered a second, more structured form of wonder when he discovered a book on Euclidean geometry. Unlike the intuitive astonishment of the compass episode, this was conscious, deductive, and luminous. He described it as a wonder “of a totally different nature” \cite{Ein49a}. The juxtaposition of these two modes—pre-verbal fascination and formal clarity—would later define Einstein’s scientific style: a creative interplay between spontaneous insight and methodical pursuit of coherence, in which the emotional power of mystery guided the rational search for lawlike unity.

Fox collapses Einstein’s intellectual development into a timeless narrative, failing to distinguish between his thoughts' scientific, personal, and philosophical dimensions as they evolved over decades. Early childhood experiences—such as his wonder at the compass or discovery of Euclidean geometry—are treated not as specific developmental episodes but as metaphysical symbols of an enduring worldview. This elides Einstein’s growing philosophical sophistication, particularly the shift from youthful empiricism to later reflections on science, religion, and metaphysics. The absence of chronological framing leads to a distortion of his intellectual trajectory. In Fox’s description of Einstein’s childhood wonder at the compass, there is a degree of philosophical retrojection: the episode is recounted not merely as a formative impression, but as if it already contained—embryonically—the conceptual depth of Einstein’s mature worldview. This risks attributing to a young child a mode of reflection that is more plausibly the product of adult retrospect, shaped by decades of scientific practice and intellectual elaboration. 

Fox's description of the wonder Einstein experienced as a child combines:

\noindent 1. A curated quote from \textit{The Ultimate Quotable Einstein} (a tertiary source) \cite{Cal}, 

\noindent 2. a secondary translation from \textit{Einstein on Einstein} (Gutfreund and Renn) \cite{GR}, 

\noindent 3. a recollected and paraphrased statement from \textit{Einstein and the Poet} (Hermanns) \cite{Her}, 

\noindent without distinguishing between these as textual genres with varying degrees of reliability and interpretive mediation. Hermann’s text, in particular, should be treated as anecdotal and unverifiable, not on par with the \textit{Autobiographical Notes} or Einstein’s own published (and unpublished) writings. By blending them seamlessly, the author erases crucial differences in textual authority.

\subsection{Significance of the Pythagorean Theorem for Einstein}

In biographies of Einstein, the “Pythagorean Theorem” is purely mathematical \cite{CPAE1}, \cite{Rei}. They do not invoke Pythagoreanism as a philosophical or mystical doctrine. Einstein’s engagement with the theorem represents an early, formative encounter with logical reasoning and the aesthetic power of Euclidean geometry, not with metaphysical ideas such as the harmony of the spheres, number mysticism, or the transmigration of souls traditionally associated with Pythagorean philosophy.

In this context, “Pythagoras” refers simply to the ancient Greek mathematician conventionally credited with the geometric result, and “Pythagorean” is used adjectivally to describe the well-known relation between the sides of a right-angled triangle. The theorem captivated the young Einstein as a mathematical challenge and source of intellectual delight. It marked an early experience of reasoning independently and discovering truth through abstraction.

Einstein's sister Maja recalls that upon successfully constructing his proof of the theorem, Albert was “overcome with great happiness.” This was not mere scholastic satisfaction, but a profound emotional response to the power of pure thought. The theorem became, for him, a symbol of intellectual emancipation and personal agency—a moment when he recognized his aptitude for abstract thinking and glimpsed the path his mind would follow.
This early mathematical achievement was not isolated. As Maja describes, it revealed to Einstein “the direction in which his talents were leading him.” His method—drawing analogies and reasoning from the similarity of triangles—foreshadowed his later heuristic style in physics. This approach, grounded in visual reasoning and conceptual analogy, became central to his mature work, including special and general relativity \cite{CPAE1}, "Albert Einstein-Beitrag für sein Lebensbild."

In this light, Pythagoras functioned not as a mystical forebear but as an emblem of mathematical clarity, deductive rigor, and formal beauty.
Thus, the Pythagorean Theorem was formative in Einstein’s intellectual biography. It represented his first serious engagement with abstract reasoning, the joy of independent discovery, and the emotional resonance of mathematical insight. While the philosophical tradition associated with Pythagoreanism had no discernible influence on Einstein’s worldview, Pythagoras' mathematical legacy held enduring personal significance for him. It symbolized for Einstein the capacity of the human mind to uncover hidden order—an early and enduring encounter with the rational intelligibility of the world. Einstein’s worldview was not mystical or metaphysical in the Pythagorean sense, and he avoided assigning any cosmic or spiritual significance to mathematical forms.

Fox’s claim that Einstein “proved to be the pinnacle of the Pythagorean tradition” \cite{Fox} is rhetorically striking, but not textually supported. Einstein's only \emph{published} reference to Pythagoreanism appears in his 1949 essay in Schilpp’s volume, where he writes that a theoretical physicist “may even appear as Platonist or Pythagorean” insofar as he emphasizes logical simplicity \cite{Ein49b}. The phrase “may even appear as” clearly signals rhetorical distance, not identification.
When Einstein recalled proving the Pythagorean theorem as a child, he described it as a formative intellectual joy, not an initiation into a mystical tradition. Elevating this moment into a rite of metaphysical passage misconstrues his reflection's tone and intent.

\subsection{Relativity: Einstein the Sage and the Angelic Cloister}

Fox presents Einstein as a figure of philosophical serenity, almost a sage cloistered in the Bern Patent Office, channeling transformative insights into physics through meditative reflection. The suggestion that Einstein’s 1905 “miracle year” emerged as a kind of sudden revelation, rather than through prolonged conceptual struggle and analytical rigor, reflects a stylized interpretation \cite{Fox}. While evocative, it risks obscuring the historically documented nature of Einstein’s creative process.

Fox claims that Einstein “established the existence of atoms,” “demonstrated the equivalence of matter and energy,” and “revolutionized our understanding of light”—transformations allegedly achieved “without doing a single experiment” \cite{Fox}. Yet the historical record is more nuanced. In his 1905 paper on mass–energy equivalence, Einstein cautiously proposed that energy loss implies mass reduction, suggesting that mass is a measure of energy content. Still, he did not yet articulate a general equivalence principle \cite{Ein5b}; broader formulations followed in 1906 and 1907 \cite{Ein6}, \cite{Ein7}. Similarly, Einstein interpreted the irregular movement of small particles suspended in a liquid as observable evidence for molecular motion—i.e., thermal motion—predicted by the kinetic theory of heat (statistical mechanics of thermal motion). Einstein had only a general awareness of the phenomenon and did not base his theory on prior experimental data, but theoretically invented Brownian motion. He independently formulated his theory in 1905 using thermodynamics and statistical mechanics, and only afterward did his theory gain empirical confirmation through experiments, especially those of Jean Perrin \cite{Stac}. 

Far from disregarding experimentation, Einstein was engaged in both conceptual and empirical concerns. In 1899, he proposed an ether-drift experiment using thermocouples and discussed ether-related phenomena in letters to Mileva Marić. He even submitted an experimental proposal to Wilhelm Wien. However, it was not realized \cite{CPAE1}, Doc. 54. Reiser and others document his active interest in experimental questions during his student years \cite{Rei}. His orientation was not mystical withdrawal, but scientifically engaged inquiry through theoretical insight and observational awareness.

\subsection{Meditations from Einstein’s Cabin}

In one of the more stylized passages of his book, Fox casts Einstein in the company of spiritual ascetics. Into this solemn procession steps Einstein the sage, sailing serenely across the Pacific, “like a monastery,” as Fox puts it, quoting \textit{The Travel Diaries of Albert Einstein} \cite{Ros}. With a few well-chosen travel diary excerpts and a late-life remark about solitude, Fox presents Einstein as a cloistered mystic in transit \cite{Fox}. It is a picturesque tableau—Einstein as a seaborne sage, adrift in metaphysical reverie—but one that will seem discordant to readers familiar with Einstein’s tone, rhetorical habits, and characteristic avoidance of grandiose metaphysical language.

Fox’s portrayal of Einstein as a quasi-monk—seeker of contemplative seclusion and practitioner of a rationalized Eastern mysticism is richly evocative yet historiographically problematic. The rhetorical strategy is clear: liken Einstein’s sea voyages to monastic retreats, elevate his solitude to a spiritual praxis, and subtly align him with the \textit{jñāna mārga}, the Hindu path of intellectual mysticism. This framing, however picturesque, distorts both the context and the content of Einstein’s writings and correspondence.

The letter to Svante Arrhenius (Jan 10, 1923, \cite{CPAE13}, Doc. 420)—quoted for its idyllic praise of sea travel—describes a kind of meditative reprieve. But Einstein’s phrase, “How conducive to thinking and working the long sea voyage is—a paradise without correspondence, visits, meetings, and other inventions of the devil,” is not the utterance of a mystic, but of a man exhausted by public obligations and grateful for silence. It is a reprieve from distraction, not a declaration of monastic renunciation.

The comparison to a “cloister” found in his letter to Niels Bohr on the same date \cite{CPAE13}, Doc. 421, is similarly tongue-in-cheek—a playful metaphor reflecting the leisurely rhythm of oceanic travel, not a statement of metaphysical doctrine. “Such a sea voyage is a splendid existence for a ponderer—it is like a cloister,” Einstein writes, wryly highlighting the conditions ideal for thought, not spiritual awakening.

This metaphor echoes his earlier reference to the Bern Patent Office as a “worldly cloister” (\textit{weltliches Kloster}). In a 1919 letter to Michele Besso \cite{CPAE9}, Doc. 207, Einstein recalls: “...into this worldly cloister, where I brooded over my most beautiful thoughts...” 
The Patent Office—often imagined as a locus of bureaucratic drudgery—was for Einstein a haven for intellectual incubation. 

In both instances, the “cloister” signifies a space of insulation from social demands and mental noise, enabling uninterrupted reasoning and theoretical development.
Thus, in Einstein’s usage, the “cloister” is a metaphor for cognitive asceticism, not spiritual transcendence. It reflects a methodological retreat from distraction, not a spiritual discipline aligned with mystical or monastic traditions. Fox’s interpretation overlays an Eastern religious inflection that exceeds what Einstein’s metaphors plausibly support. 

Einstein’s solitude was methodological, not metaphysical. His retreat was not a sādhanā, but an escape from administrative demands—a quest for clarity, not cosmic communion. Recasting such reflections as spiritual invocations misreads Einstein’s tone and misconstrues the functional role solitude played in his intellectual life. Far from whispering divine truths in isolation, Einstein used such spaces to brood over physical problems, formulate thought experiments, and develop conceptual breakthroughs. Fox’s interpretive leap turns an ironic metaphor into a theological statement, thereby misaligning the secular tenor of Einstein’s remarks with a spiritualized narrative arc.

Fox writes that Einstein always pursued what he saw as the core of the spiritual life, insisting that “I always loved solitude” \cite{Fox}.
This oft-cited phrase \cite{Cal} comes from a 1952 letter, written when Einstein was in his seventies, long after the period of his major scientific breakthroughs. It is a personal and reflective remark, embedded in a private context. Yet some commentators have extrapolated from this modest statement elaborate philosophical and psychological narratives. In certain portrayals, Einstein becomes a contemplative sage, and his solitude is interpreted as a sign of quasi-mystical withdrawal. In others, it serves as a retroactive basis for speculative psychological diagnoses, such as Asperger’s syndrome, reframing his introspective temperament as a clinical symptom.

In Fox’s rendering, Einstein’s inclination toward solitude is woven into a larger metaphysical narrative, casting him as a cloistered sage who sought intellectual transcendence through rational mysticism. In this framing, intellectual withdrawal becomes cosmic attunement; solitude becomes spiritual vocation.
The metaphor of the monastic retreat is extended across multiple domains of Einstein’s life, from his sea voyages and patent office work to his late studies, creating an image that departs significantly from Einstein’s self-understanding.

\section{Unified Field Theory} \label{4}

\subsection{Mystical Overreach in the Interpretation}

Fox’s \textit{I Am a Part of Infinity} attempts to recast Einstein’s scientific worldview—particularly his commitment to unified field theory—within a spiritual metaphysics grounded in nonduality, cosmic interconnection, and ego transcendence. While Fox's narrative is often eloquent with respect to Einstein’s spiritual language and ethical universalism, his interpretation \emph{blurs} critical distinctions between Einstein’s technical physics and the metaphysical or mystical traditions he seeks to align it with. A close reading of Einstein’s actual unified field theory writings, including his 1923–1929 papers, reveals a substantial misreading on several fronts.

\subsection{Einstein's Determinism is Mathematical}

Fox frequently presents Einstein's determinism as a quasi-spiritual belief in cosmic interdependence—a metaphysical tapestry in which individuality dissolves and human agency is revealed as illusory. He interprets Einstein’s view of causality as an extension of Taoist and Vedantic thought, suggesting that Einstein’s critique of naive cause-and-effect thinking reflects a mystical worldview akin to Spinoza’s sub specie aeternitatis or Advaita Vedanta’s Brahmanic unity. However, this conflates Einstein’s methodological commitment to deterministic field equations with a metaphysical doctrine of inter-being. In his unified field theory writings, such as the 1923 paper “Bietet die Feldtheorie Möglichkeiten für die Lösung des Quantenproblems?” \cite{CPAE14}, Doc. 170—Einstein consistently frames determinism in terms of mathematical overdetermination of partial differential equations. There is no trace in these technical discussions of a desire to transcend individuality or merge consciousness with cosmic unity. The determinism Einstein defends is a feature of field-based continuity and causal structure within a four-dimensional spacetime, not an ontological commitment to spiritual unity.

\subsection{Overinterpretation}

One of Fox’s core claims is that Einstein’s effort to eliminate the dualism between field and matter signifies a philosophical move toward nonduality, equating this with Spinoza’s monism and Eastern notions of unity. However, Einstein’s rejection of the field-matter dichotomy is not motivated by a metaphysical vision of oneness; it arises from specific physical considerations. For Einstein, describing particles as singularities or localized concentrations of field energy was an attempt to reconcile general relativity with the phenomena of quantum mechanics and electrodynamics, not to resolve spiritual alienation. In his papers and manuscripts on unified field theory, Einstein is preoccupied with finding an overdetermined system of covariant field equations that will predict particle-like solutions (e.g., electron configurations) without invoking quantum discontinuities. He is explicit that his goal is to derive singularities (representing electrons) as consequences of the field equations themselves—solutions $L(m, \varepsilon)$ representing mass and charge, not to offer a cosmological spirituality of unity \cite{CPAE14}, Doc. 170.

Thus, when Fox reads Einstein—more precisely, Infeld’s formulation, since Einstein did not author but merely edited \textit{The Evolution of Physics} (see section \ref{4})—that “the division into matter and field is something artificial” \cite{IE} as evidence of metaphysical nonduality, he misconstrues Einstein’s actual meaning. This statement refers to physical continuity, not metaphysical identity. For Einstein, the field is the fundamental physical entity, and matter is a specific energetic concentration within it, not an illusion, and indeed not a metaphor for spiritual self-transcendence.

\subsection{Misappropriating the Field}

Fox’s most serious overreach lies in his repeated portrayal of Einstein’s field theory as a form of spiritual realization. He writes, for example, that “field being the only reality” (again quoting Infeld! \cite{IE}) is Einstein’s way of expressing that “all were woven from a single shimmering fabric.” However, Einstein and Infeld’s assertion that the field is the sole ontologically fundamental entity is a technical claim, grounded in Einstein's rejection of the probabilistic framework of quantum theory and his quest for a continuous, deterministic field-based description of physical reality. It should not be conflated with spiritual or poetic metaphors of divine unity. Nowhere in Einstein’s unified field theory writings does he suggest that the comprehension of this physical theory carries existential or transformative implications for the self. The transition to a pure field ontology is a methodological objective, not a mystical epiphany.

Moreover, Einstein’s continued insistence on rigorous mathematical consistency, general covariance, and the inclusion of empirical singularities in the field equations sharply distinguishes his program from any intuitive or contemplative tradition. The scientific style of the field equations—nonlinear, overdetermined systems of second-order tensor equations—leaves no space for the intuitive gnosis that Fox projects onto Einstein’s program.

\subsection{A Misreading}

Fox’s invocation of Spinoza, Schopenhauer, the Upanishads, and Taoism results in a spiritual synthesis that, while imaginative, lacks fidelity to Einstein’s actual scientific goals. His narrative performs a poetic syncretism in which Einstein’s technical vocabulary is repeatedly repurposed to support a metaphysical worldview Einstein never endorsed. For example, Fox states that Einstein’s determinism reflects “an uplifting intuition of being interwoven with all things.” Still, in the 1923 field theory paper \cite{CPAE14}, Doc. 170, determinism is discussed as a requirement for constructing a temporally consistent field theory, not as an existential intuition. 

Fox’s portrayal of Einstein as a spiritual figure advancing a nondual metaphysics through field theory is an interpretive construction rather than a historically grounded analysis. Einstein’s writings on unified field theory do not support the notion that he viewed the field as a mystical principle or equated scientific unification with spiritual transcendence. His project was deeply rationalist, not mystical; his commitment was to causal, mathematical coherence, not ontological dissolution of the self. While Einstein’s cosmic religious feeling is well-documented and undoubtedly sincere, Fox conflates this feeling with the entirely distinct aims of his scientific work. In doing so, he produces a moving but ultimately misleading image of Einstein—a composite of scientist and sage that says more about the modern hunger for synthesis than it does about Einstein himself.

\section{Lack of Source Discrimination} \label{5}

\subsection{Primary vs. Secondary vs. Tertiary}

A central shortcoming of Fox’s thesis lies in the methodological looseness with which it is constructed. The argument rests on an insufficiently discriminating use of sources, often drawing from anthologies such as \textit{The Ultimate Quotable Einstein} \cite{Cal}, and \textit{Albert Einstein: The Human Side} \cite{DH} without careful contextualization or source criticism. This weakens the historical and philosophical foundation upon which his spiritual reading of Einstein is built.
Alice Calaprice’s \textit{The Ultimate Quotable Einstein} \cite{Cal} is a widely used reference that organizes Einstein’s remarks thematically, often detaching them from their original context, language, and argumentative framework. Even when specific quotations cannot be cross-referenced because the source is unavailable, historians and philosophers draw on broader contextual knowledge—Einstein’s corpus, chronology, and intellectual milieu—to assess a quote’s reliability, tone, and interpretive weight. The problem here lies not in citation per se, but in the uncritical elevation of isolated remarks to thesis-defining status.

This pattern extends to other mediated sources such as Denis Brian’s \textit{Einstein: A Life} \cite{Bria}, which blends secondary quotations with narrative commentary. Although such works are accessible to broader audiences, they often compress, paraphrase, translate, and editorially frame into a unified voice, thereby obscuring the complex mediation between Einstein’s statements and their present form.

Rather than engaging with Einstein’s original \textit{Autobiographical Notes}, Fox cites \textit{Einstein on Einstein} by Hanoch Gutfreund and Jürgen Renn no fewer than fifteen times—an extensive reliance on a secondary source that further distances his thesis from Einstein’s framing of his intellectual development. While \textit{Einstein on Einstein} offers valuable interpretive framing, it does not replace direct engagement with Einstein’s original \textit{Autobiographical Notes}. Scholarly convention continues to privilege documents such as the \textit{Autobiographical Notes} in Paul Arthur Schilpp’s \textit{Albert Einstein: Philosopher–Scientist} \cite{Schi}—because it includes Einstein’s original \textit{Autobiographisches} in German—as the most authoritative articulation of his philosophical outlook. Secondary commentary should illuminate such primary sources, not overshadow or supplant them.

Another problematic instance is the recurrent use of William Hermanns’ \textit{Einstein and the Poet} \cite{Her}, a book composed decades after the conversations it purports to recount. It relies on retrospective accounts that often exhibit a stylized, literary tone blending memory, interpretation, and embellishment. Without corroborating records, such sources must be treated as anecdotal, not cited as reliable stand-ins for Einstein’s documented writings. Hermanns' Conversations with Einstein captures a striking—and surreal—juxtaposition between Einstein's lofty, mathematically grounded worldview and the mystical and religious perceptions projected onto him by others. Hermanns' account includes poetic embellishment, descriptive flourishes, and anecdotal color, particularly in the dialogue's tone, structure, and dramatic contrasts. These literary devices are typical of memoiristic reconstruction rather than verbatim transcription. They signal an intention to dramatize, not simply record.

Fox’s narrative at the beginning of his book—describing his time living with Tibetan monks in the Himalayas, his training in meditation, and his longing for spiritual seclusion \cite{Fox}—strongly predisposes him to resonate with mystical or spiritual portrayals of scientific individuals like Einstein. This background helps explain why Hermanns’ Conversations with Einstein would resonate profoundly with him. Fox and Hermanns both frame Einstein through the lens of spiritual yearning, not just scientific inquiry.

In reading Fox’s autobiographical reference to his time living with Tibetan monks, I was reminded of Hermanns’ account of Einstein listening to stories about Tibetan monks. 
In a brief exchange reported by Hermanns, Einstein plays a familiar role \cite{Her}: the reluctant prophet to whom the spiritually inclined and the mystically inspired come seeking validation. The scene is theatrical, almost allegorical. A Christian minister, fervent and pale, and an Indian monk are invoked in the same breath, each bearing claims of supernatural phenomena: Tibetan monks walking through walls, remote spiritual healing. In this context, Einstein becomes a symbol of transcendent truth not because of his scientific contributions but because of the symbolic aura surrounding him.
Einstein, pipe in hand and retreating into his chair with a faint smile, performs his quiet skepticism not by direct refutation, but through contrast. When he finally responds, his tone is a model of Socratic irony. He does not mock; rather, he sidesteps the mystical claims by invoking the names of Euclid, Leibniz, Gauss, and Riemann—not merely as intellectual ancestors but as representatives of a worldview rooted in rational structure and mathematical beauty. 

While Hermanns’ account may reflect the essence of Einstein’s views, it almost certainly involves literary shaping. The dialogue serves a rhetorical function—Einstein as the calm, enlightened rationalist standing against mystical credulity—and should not be taken as a verbatim historical transcript.

A further example of problematic source use is the frequent citation of the so-called "Socratic Dialogue" between Max Planck, Einstein, and James Murphy, presented as an epilogue in Planck’s \textit{Where Is Science Going?}. In this dialogue, Einstein is given a prominent voice in a stylized exchange on science, philosophy, and metaphysics. However, the editorial framing of the text makes clear that it cannot be regarded as a primary source. Planck explicitly states that “the following is an abridgment of stenographic reports made by an attendant secretary during various conversations” \cite{Pla} This statement immediately signals the text’s indirect nature: it is a reconstruction, filtered first through stenographic notes, then shaped by Planck himself—who transformed fragmentary recollections into a literary dialogue modeled on Platonic conventions—and finally mediated by the translator, James Murphy, in rendering the text from German into English.

This undifferentiated use of materials results in a blurred evidentiary hierarchy. Eyewitness recollections are presented alongside carefully composed philosophical statements, paraphrases are cited as if they were verbatim, and translated anthologies are treated as equivalent to original manuscripts. This flattening of source levels undermines the narrative's interpretive integrity.

Source discrimination is not optional but essential in scholarly work, particularly when reconstructing a thinker as complex and historically situated as Einstein. Without attention to textual provenance, historical context, and the layers of interpretive mediation, Einstein’s intellectual and philosophical portrait becomes oversimplified and potentially distorted. Careful differentiation among primary, secondary, and tertiary sources is foundational to preserving historical accuracy and conceptual clarity.

\subsection{Einstein and Solovine: Historical Appreciation, Not Metaphysical Commitment}

An illustrative example of interpretive overreach appears in Fox’s discussion of Einstein’s marginal annotations in Maurice Solovine’s Démocrite. Fox claims that “Einstein very rarely made markings in his books, but Solovine’s Democritus is graced with many of Einstein’s handwritten highlights,” suggesting that Einstein embraces the Pythagorean tradition \cite{Fox}. While Einstein’s annotations indicate interest, they do not justify philosophical alignment or ideological transformation claims. Unless they contain explicit affirmations of agreement or statements of personal identification, Marginalia are insufficient grounds for asserting that Einstein adopted a Pythagorean worldview.

The fact that Solovine mentions, on the third page, that Democritus was “a zealous follower of the Pythagoreans” cannot serve as decisive evidence, especially without knowing whether Einstein marked that specific passage. Even if he did, such reading does not entail endorsement. Readers engage with texts critically, selectively, and sometimes skeptically. 
\emph{Mere ownership of a book in one’s personal library, or the presence of annotations, does not imply wholesale acceptance of its philosophical content. Books may be received as gifts—often from colleagues or friends—and accepted out of courtesy rather than conviction; their presence on the shelf may reflect personal relationships or intellectual curiosity rather than philosophical endorsement.} Unless Fox has systematically reviewed all of Einstein’s handwritten comments in Démocrite and found explicit statements confirming that Einstein regarded himself as an heir to the Pythagorean tradition, his conclusion remains speculative and unsubstantiated.

Fox draws attention to Einstein’s copy of the Tao Te Ching in the Albert Einstein Archives—remarking that it contains handwritten markings “alongside almost a quarter of all the passages,” and noting that this constitutes more annotation than in nearly any other volume in Einstein’s personal library \cite{Fox}. However, he does not examine or cite the actual content of these annotations. His inference that Einstein “clearly studied [it] carefully” \cite{Fox} is thus based solely on the presence of marginalia, not on any close reading of the remarks themselves.
Without quoting Einstein’s marginalia, Fox cannot justifiably claim that the Tao Te Ching held special spiritual significance for Einstein. As a counterexample, it is conceivable that Einstein found the text obscure, or even fundamentally incompatible with his scientific worldview, and marked it accordingly.

In this respect, Fox’s failure to investigate the annotations more deeply undercuts the evidentiary value of his observation. If Einstein had written, for instance, marginal notes expressing disbelief, irony, or even private irreverence—what Fox characterizes as reverent engagement would instead be reframed as analytical scrutiny or detached curiosity. Fox’s claim remains speculative and vulnerable to alternative interpretations without such content analysis.

Fox cites a letter Einstein wrote to Maurice Solovine in 1930 telling him he had been “elated” after reading it \cite{Fox}. This letter from Einstein to Solovine, dated 4 March 1930, was later published in \emph{Lettres à Maurice Solovine}. I will provide below an excerpt from this letter by Einstein that is not quoted in Fox’s book but is highly relevant to the themes under discussion. In this excerpt from the letter, Einstein compliments Solovine's comparison of Democritus and Pythagoras by noting that \cite{Sol} (my translation from German to English): 

\begin{quote}
What gave me the greatest pleasure was your exposition of the development of Greek thought, particularly your presentation of Democritus' doctrine compared to that of Pythagoras. I found it not only successful but also enlightening. One sees that in the case of the atomists, there is a parallelism with modern mechanistic ideas (atoms and the void). On the other hand, the Pythagorean theory of numbers, with its tendency toward harmonics, reminds us more of contemporary field theory. The Greek thinkers were truly remarkable in their penetration and originality.    
\end{quote}

Einstein notes that Democritean atomism resembles modern mechanistic theories (atoms and void), while Pythagorean numerology and harmonic theory evoke modern field theory (i.e., a mathematical, continuous view of nature). He uses phrases such as “reminds us more of contemporary field theory” and “parallelism with modern mechanistic ideas”—indicative of conceptual analogies rather than philosophical commitments.

These historical resonances appealed to Einstein as a physicist attuned to formal structure and theoretical elegance. While Solovine is credited with the interpretive framework, Einstein affirms its value as intellectually stimulating and heuristically insightful.

However, this correspondence does not imply that Einstein identified with Pythagoreanism or subscribed to its metaphysical or mystical doctrines. As far as current sources attest, neither Einstein’s published writings nor his surviving private correspondence offers any endorsement of Pythagorean philosophy as a substantive worldview. Though he acknowledged affinities between certain ancient ideas and developments in modern physics, such remarks were illustrative, not doctrinal. Unless new materials surface, there is no textual basis for treating Einstein as a Pythagorean in metaphysical or spiritual terms.

\subsection{Deficiencies in interpreting Sources}

Further complications arise from the use of \textit{The Evolution of Physics} \cite{IE}, a popular science volume co-authored with Leopold Infeld, to support claims about Einstein’s alleged “Pythagorean faith.” While Fox treats this work as evidence of metaphysical conviction, the source is more complex. Infeld drafted much of the text in accessible English prose, and Einstein’s role was primarily editorial \cite{Inf}, \cite{Sta}. Such remarks caution against interpreting the book as an unfiltered articulation of Einstein’s philosophical views.

Fox presents Einstein’s approach to theory construction as one characterized by epistemic humility and an acceptance of error as intrinsic to scientific creativity. Drawing on metaphors such as mountain climbing and fragmented consciousness, Fox suggests that Einstein viewed intellectual development as a gradual ascent in which false starts, detours, and partial perspectives are not discarded but incorporated into broader, more comprehensive visions. Theory-building, in this framing, is evolutionary rather than revolutionary—a cumulative expansion of understanding rather than the replacement of old structures.

However, Fox’s account rests largely on quotations from Einstein’s \emph{Autobiographical Notes} and \emph{The Evolution of Physics}, without attending to the complex provenance of these texts. For instance, the metaphor of mountain climbing is cited from \emph{The Evolution of Physics}, which blurs the distinction between Einstein’s personal philosophical commitments and collaborative, pedagogical narrative.
Similarly, Fox cites from Gutfreund and Renn’s translated edition of the \emph{Autobiographical Notes}, rather than from the original version published in Schilpp’s \emph{Albert Einstein: Philosopher–Scientist}, where the tone and framing are arguably more circumspect. This reliance on mediated and collaborative sources raises questions about the weight Fox assigns to them in reconstructing Einstein’s philosophical stance, especially when more authoritative primary texts remain available but unexamined.

Einstein’s attitude toward error, particularly as documented in his correspondence, provides a revealing counterexample to the stylized, spiritualized portrait presented by Fox. In a 1915 letter to Arnold Sommerfeld, written shortly after formulating the final field equations of general relativity, Einstein remarked with characteristic irony that he had “immortalized” his earlier mistakes in the Academy papers \cite{CPAE8}, Doc. 153.\footnote{“Die letzten Irrtümer in diesem Kampfe habe ich leider in den Akademie-Arbeiten, die ich Ihnen bald senden kann, verewigt.”} 
Notably, Einstein misspelled verewigt as verevigt, adding a layer of unintentional self-irony. 

This acknowledgment of fallibility—accompanied by a typographical error in the very word verewigt—epitomizes Einstein’s modest and unsentimental approach to scientific development. His conceptual progress was not cloaked in mythic ascent or metaphysical yearning, but grounded in hard-won insight, self-correction, and disciplined rethinking. He did not treat errors as “meandering” \cite{IE} detours in a transcendental ascent to illumination, but as intrinsic to the rational reconstruction of physical theory. Rather than obscuring his missteps, he often foregrounded them, reinforcing his identity as a thinker more concerned with intellectual honesty than philosophical or spiritual aura. This self-aware, frequently ironic stance stands in tension with Fox’s portrayal of Einstein as a spiritual exemplar and metaphysical visionary.

Relying on paraphrased citations, imprecise attributions, and selective quotation contributes to what might be called a quasi-hagiographic narrative of Einstein: one that elevates him into a figure of universal wisdom while erasing the historical tensions, intellectual developments, and rhetorical ironies that animate his actual writings. The failure to consistently engage with Einstein’s original writings and to critically assess the provenance, context, and rhetorical function of widely circulated quotations results in an interpretive framework that rests on unstable ground. Anecdotes, poetic renderings, and posthumous recollections cannot substitute for disciplined source analysis. Without careful attention to source hierarchy, translation accuracy, and documentary context, historical portraiture becomes vulnerable to distortion, regardless of its narrative appeal.

\subsection{Schizophrenia, Modern Malaise, and Methodological Inconsistency}

In scholarly practice, secondary and paraphrased sources must be triangulated with Einstein’s primary corpus—his scientific works, manuscripts, letters, and philosophical reflections. These texts, especially as collected in the \textit{Collected Papers of Albert Einstein}, provide the necessary historical, linguistic, and conceptual scaffolding for responsible interpretation. Notably, Fox makes minimal use of the \textit{Collected Papers of Albert Einstein}, citing them only sparingly—approximately ten times throughout the book, predominantly from volume 15. Rather than engaging systematically with Einstein’s primary writings, Fox relies heavily on secondary sources, popular biographies, and interpretive commentary. This preference limits the historical fidelity of his account and raises concerns about the depth of his engagement with Einstein’s scientific and philosophical expressions.

One of the few moments in Fox’s book where he engages substantively with volume 15 of the \textit{Collected Papers of Albert Einstein} is in his discussion of Einstein’s correspondence with his son Eduard, who suffered from schizophrenia. Here, Fox shifts from interpretive speculation to close quotation, citing Einstein’s words with a degree of fidelity and contextual attention that is largely absent elsewhere in the book. This sudden methodological rigor is striking. In the case of schizophrenia, Fox not only turns to primary sources but also constructs a literary-philosophical reading of the letters that seeks to position Eduard’s despair as emblematic of the “modern mind’s malaise.”

Although Fox does not explicitly identify schizophrenia as the malaise of modernity, he mobilizes Eduard’s worldview, characterized by nihilism, disillusionment, and existential pessimism, as a symbolic condensation of modern alienation. Eduard’s remarks, such as his claim that human history is “completely insignificant” \cite{CPAE15}, Doc. 274, or that there is “a desperately small difference between a genius and an idiot” \cite{CPAE15}, Doc. 414, are quoted not as psychiatric symptoms but as profound, if troubling, philosophical insights. In framing them this way, Fox transforms a clinical and familial tragedy into a cultural allegory. Schizophrenia implicitly becomes a narrative device for staging the metaphysical uncertainties of the twentieth century, rather than a psychological condition warranting biographical sensitivity \cite{Fox}.

Fox’s treatment of Einstein’s reaction to his son’s illness follows a similarly stylized trajectory. He quotes Einstein’s response—“One must not take oneself too seriously. For one is a critter that barely became two-legged via the ape...” \cite{CPAE15}, Doc. 257—as evidence of Einstein’s cosmic humility and stoic detachment \cite{Fox}. But he offers no exploration of Einstein’s emotional burden, no analysis of the father’s possible guilt, avoidance, or philosophical reckoning with his son’s suffering. Instead, Einstein becomes a foil for Eduard’s despair: the sage-scientist who answers existential dread not with anguish, but with humorous fatalism and philosophical distance.

This literary treatment is notable not only for its symbolic overtones, but also for its inconsistency. Only here, Fox quotes extensively from the \textit{Collected Papers of Albert Einstein}. The rest of the book demonstrates a marked preference for secondary, often anecdotal, sources. It tends to reframe Einstein’s thought through the lens of Hinduism, Buddhism, Taoism, and other non-Western metaphysical systems. Yet when dealing with the emotionally charged and philosophically complex topic of schizophrenia, Fox refrains from importing the Eastern frameworks that otherwise saturate his interpretation of Einstein’s spiritual worldview. There is no invocation of Brahmanic unity, jñāna mārga, moksha, Taoist non-action, or Buddhist non-self. Instead, Fox allows the letters to resonate in a purely existential register marked by cosmic humility, finitude, and alienation.

This asymmetry reflects a broader limitation of the book: Fox is not a historian of science or a physicist, which becomes evident in his treatment of Einstein’s scientific work. His framing tends toward symbolic or spiritual reinterpretation rather than historically grounded contextualization. The result is that while Fox can analyze a letter about schizophrenia with textual care and psychological sensitivity, he does not bring the same methodological precision to Einstein’s published scientific papers or philosophical reflections, which are instead reimagined through the lens of spiritual syncretism.

\subsection{Chronological Flattening}

A further methodological issue in Fox’s account is the tendency to collapse distinct phases of Einstein’s intellectual development into a temporally flattened narrative. Texts and quotations from widely separated decades—such as the 1930 essay "What I Believe" \cite{Ein30} and retrospective recollections from the 1940s and 1950s—are cited without regard for the intervening transformations in Einstein’s philosophical outlook. This chronological conflation presents Einstein as a static thinker whose worldview remained fully formed over decades, rather than a historically situated figure whose ideas evolved in response to changing scientific, political, and personal contexts.

This flattening obscures crucial transitions in Einstein’s intellectual and biographical trajectory. In 1930, Einstein was still based in Berlin, embedded in the vibrant public life of the Weimar Republic and deeply engaged with European debates about science, religion, and pacifism. By 1933, he had emigrated to the United States in response to the rise of Nazism. From 1940 until he died in 1955, he lived and worked at the Institute for Advanced Study in Princeton, increasingly focused on ethical responsibility, Zionism, nuclear disarmament, and the philosophical foundations of physics. To treat Einstein’s later writings as continuous with his interwar reflections, without acknowledging these profound dislocations, is to overlook the dynamism and responsiveness of his intellectual life.

Fox also retrojects Einstein’s later emphasis on mathematical elegance into his early period. A 1931 statement about being guided not "by the pressure from behind of experimental facts", but drawn by “mathematical simplicity” \cite{Isa}, \cite{Fox} is used to frame his 1905 insights, but this is anachronistic. In his \textit{Autobiographical Notes}, Einstein confesses that he found mathematics somewhat dispensable as a student, only embracing its centrality during his general relativity work a decade later \cite{Ein49a}. At the time of the 1905 papers, his thinking was shaped more by physical intuition and conceptual clarity than by aesthetic mathematical criteria \cite{Pais}.

\subsection{Relying on Translated Anthologies}

In addition to chronological compression, Fox’s reliance on English-language anthologies introduces further interpretive distortion. Many of Einstein’s key philosophical texts were written in German, and his prose's tonal and conceptual precision is often attenuated in translation. 

A case in point is Fox's frequent use of "awe" to characterize Einstein’s attitude toward the cosmos, science, and existence. Let us translate this term into German, in a context that aligns with Einstein’s native language and usage. “Ehrfurcht” is the appropriate equivalent. It is typically translated as “awe” or “reverence.” 

However, consider, for example, a 1911 letter to Marie Curie, in which Einstein writes \cite{CPAE8}/\cite{CPAE5}, Doc. 312a: “Aber ich bin überzeugt, dass Sie diesen Pöbel stets gleich verachten, ob er Ehrfurcht heuchelt oder seine Sensationslust durch Sie zu stillen sucht!” 
Here, "Ehrfurcht" is used ironically, referring to the public’s feigned reverence—a performative display masking base motives. Far from expressing spiritual elevation, the term becomes a rhetorical device to mock hypocritical public sentiment. By contrast, in more earnest contexts—such as his writings on “cosmic religious feeling”—Einstein employs "Ehrfurcht" to evoke epistemic humility in the face of nature’s rational structure. Even so, his tone remains philosophical and non-devotional in these moments, resisting the sanctifying connotations often introduced in English translation.

The conceptual point here is critical: Translation choices can subtly inflect Einstein’s meaning, amplifying moral or spiritual undertones he did not intend. A historically and philosophically grounded reading of Einstein requires attention to textual provenance and the idiomatic texture and rhetorical purpose of his original German expressions.

By compressing chronological distinctions and relying heavily on translated anthologies, Fox's narrative loses sight of these essential nuances. The result is a stylized and largely decontextualized portrait that effaces Einstein’s rhetorical range, evolving commitments, and adaptive style. Rather than tracing the transformations of a thinker engaged with some of the most turbulent developments of the twentieth century, the account presents a homogenized persona, removed from historical complexity and flattened into a symbol of perennial spiritual insight.

\section{Conclusion}

Kieran Fox’s central thesis—that Einstein’s worldview constitutes a coherent spiritual journey culminating in a pantheistic cosmology grounded in wonder, unity, and nonduality—offers a provocative reframing of Einstein as not merely a scientist, but a modern sage. This interpretation draws on Einstein’s expressions of awe and his references to Spinoza, Schopenhauer, and Eastern philosophy to construct a narrative of spiritual development that mirrors the stages of mystical ascent in various traditions.

Yet upon closer examination, this thesis rests on an unstable foundation. It merges historically disparate influences and symbolic resonances into a unified spiritual arc that Einstein himself neither articulated nor endorsed. 

More significantly, the thesis relies on methodological practices undermining its scholarly credibility. By drawing indiscriminately from heterogeneous sources—Einstein’s scientific writings, anecdotal memoirs, third-party paraphrases, and philosophically divergent traditions—Fox erases the internal tensions and historical specificity of Einstein’s intellectual development. The result is not a reconstruction of Einstein’s philosophical worldview but a retrospective synthesis that imposes coherence where there was complexity, and spiritual teleology where there was dialectical struggle.

Fox’s project invites reflection on the broader responsibility of intellectual history. To recover Einstein’s cosmic sensibility is not to reduce it to pantheistic metaphysics or Eastern syncretism, but to situate it within his scientific worldview's cognitive, ethical, and methodological commitments. Einstein’s remarks on wonder, determinism, and reverence for nature gain their meaning not from mystical speculation, but from a disciplined naturalism informed by empirical science, mathematical clarity, and philosophical restraint.

In summary, if Einstein is to be remembered as a sage, let it be with chalk in hand before a blackboard, a pipe between his lips, and perhaps a stray violin note drifting slightly off-key—not ascending to the heavens, but remaining grounded, quietly striving to make sense of what little nature permits us to understand.

\end{document}